\documentclass[12pt]{article}
\usepackage{graphics,epsfig}
\usepackage[english]{babel}
\begin{document}


\begingroup

\raisebox{0.5cm}[0cm][0cm] {
\begin{tabular*}{\hsize}{@{\hspace*{5mm}}ll@{\extracolsep{\fill}}r@{}}
\begin{minipage}[t]{3cm}
\vglue.5cm
\end{minipage}
&
\begin{minipage}[t]{7cm}
\vglue.5cm
\end{minipage}
&
\begin{minipage}[t]{7cm}

\end{minipage}
\end{tabular*}
}

\begin{center}

{\Large{\bf Polarized gamma-beams on the base of excited crystal}}

\vspace{1.cm}
                {\bf A.Aganyants}

\setlength{\parskip}{0mm}
\small

\vspace{1.cm}
              {\bf Yerevan Physics Institute, Armenia}
\end{center}

\vspace{1.cm}

\begin{abstract}

  Possibilities of creating high energy polarized photon beams at the end of
   bremsstrahlung spectrum and a method for analysis of degree of beam
    polarization  are considered in this paper.

\end{abstract}

\newpage
  Investigations of interactions of intense relativistic electron beam
   with single crystal have resulted in observing electron beam intensity
    effect in radiation \cite{ref1}, i.e. non-linear anomalous radiation.
    One can think nature of that is connected with synchronization
    of atomic Coulomb-fields, which means creating strong field for
    ultra-relativistic electrons passing through the oriented crystal.
        This effect is manifested stronger on the edge of crystal.
   Apparently, other electromagnetic process-pair production
   is modified also. Main causes for such phenomenon is exciting
   of the crystalline medium and spatial asymmetry on the crystal edge.
  We propose to create polarized photon beams at the end of bremsstrahlung
   spectrum by means of both different ways on the base of obtained results
   \cite{ref1}:\\
  1. Electrons entering this strong field emits intense beam of
   photons
   with high degree of polarization \cite{ref1}, as one
   is synchrotron -like radiation \cite{ref2}.
   According to \cite{ref2} increase of field strength will result in displacement
   of photon peak to ${\omega}_{c} \sim {\gamma}^3 /R$
   (where $\gamma$-Lorents-factor of electron,
   $R$- curvature radius) nearly to electron primary energy.
   Such field on the crystal edge can be received if to use narrow
   very intense electron beam. However at working with tagged photon the
   high intensity electron beam can not be used. In this case $X$-ray
   tube or laser can excite crystal. \\
   2. High energy unpolarized photon beam entering mentioned strong
   field of oriented crystal will be converted to electron-positron
   pairs with different cross-sections in depending on orientation of crystalline
   planes relative to direction of linear photon polarization.
   One can receive then linear polarized photon beam using bremsstrahlung
   at the end of its spectrum. However in contrast to known method of
    selective absorption the degree of photon beam polarization can be
     higher in case of excited crystal. Let us consider such a possibility
     more in detail. Production of electron- positron pair in field of
      atomic nucleus is process reverse to bremsstrahlung, as it is
      other cross-channel of bremsstrahlung process. It means effect
      intensity must be manifested also at pair production in oriented
      crystal. However in this process minimal longitudinal momentum
       transfer to nucleus $\delta$ = 2/${\omega}$ \cite{ref3} (where $\omega$
        - photon energy) is
        more than in bremsstrahlung Therefore the more energy
        of photon the more cross-section of pair production.
    So arising large difference of pair production cross-sections in excited
   oriented crystal in depending on the direction of linear photon
  polarization with respect to the crystalline  planes can result
  in desired results. In paper \cite{ref4} was shown the growth of radiation
   cross-sections with electron beam intensity. One can expect appropriate
    growth at pair production. Then one can carry out experiment \cite{ref5}
    with more thin corundum crystal, varying $\exp({-{\sigma}_{\perp}\cdot{t}})$,
     responsible for photon absorption (where $ {\sigma}_{\perp}$- growing
      cross-section, $t$- crystal thickness) to ensure high degree of the photon beam
     polarization and gamma beam intensity.

  One can use mentioned phenomenon for increasing analyzing power of
   linear polarized gamma rays using single crystal by analogy of
    Cabibo method \cite{ref6} However passing of high-energy gamma rays
    through crystal do not accompany with high excitation of crystal,
     as it takes place in case of electrons. That is way $X$-ray tube or
      laser is necessary here.

   Artificial birefringence arises in crystal in fact when pair production
  cross-section depends on orientation of linear photon polarization with
   respect to crystalline planes. Then excited single crystal of
  appropriate thickness can be used also as the high-energy analogy
  (showed by Cabibo \cite{ref7}) of a quarter wavelength plate to convert
  linear polarized photon beam into circular one.

  All subjects touched on this paper present an interest for spin physics
   therefore they are needed in further experimental investigations on
   appropriate accelerators. 

The work is supported by CRDF AP2-2305-YE-02.


\end{document}